\documentstyle[aps,prl,psfig,floats]{revtex}

\begin{document}

\draft \voffset 1cm

\twocolumn[\hsize\textwidth\columnwidth\hsize\csname %
@twocolumnfalse\endcsname

\title{Exponential Temperature Dependence of Penetration Depth in Single Crystal MgB$_2$}
\author{F.~Manzano, A.~Carrington, and N.E.~Hussey}
\address{H.~H.~Wills Physics Laboratory, University of Bristol, Bristol, BS8 1TL, England.}
\author{S. Lee, A. Yamamoto, and S. Tajima}
\address{Superconductivity Research Laboratory, ISTEC, Tokyo 135-0062, Japan}

\maketitle

\begin{abstract}
The temperature dependence of the London penetration depth, $\lambda(T)$, was measured in both single crystal and
polycrystalline MgB$_2$ samples by a high-resolution, radio frequency technique. A clear exponential temperature
dependence of $\lambda(T)$ was observed at low temperature, indicating $s$-wave pairing. A BCS fit to the lowest
temperature data gives an in-plane energy gap $\Delta$ of $30\pm2$ K ($2\Delta/T_c=1.5\pm0.1$), which is significantly
smaller than the standard BCS weak coupling value of 3.5. We find that the data are best described by a two-gap model.
\end{abstract}

 ]

The recent discovery \cite{nagamatsu} of superconductivity at 39~K in the binary compound MgB$_2$ has sparked a large
number of investigations into its physical properties.  A crucial question is whether its high $T_c$ can be explained
by a phonon mediated pairing interaction within the usual BCS-Eliashberg framework. A first step in answering this
question is to determine the symmetry of the superconducting order parameter and the nature of the low energy
excitations.

Magnetic penetration depth ($\lambda$) measurements are a powerful probe of the low energy excitations.  As
$\lambda(0)>1000$~\AA~in MgB$_2$, penetration depth measurements probe relatively large distances and are far less
sensitive to surface quality than, for instance, tunneling, which is sensitive to disorder on the scale of the
coherence length ($\xi\sim$ 50~\AA). We therefore expect the results to be representative of the bulk.

There have been several recent reports of measurements of $\lambda(T)$ for  MgB$_2$  by ac-susceptibility
\cite{panagopoulos,chen}, muon spin rotation \cite{panagopoulos} and optical conductivity \cite{pronin} techniques.
These authors conclude that $\lambda(T)$ follows a power law dependence [$\lambda(T)\sim T^2$
\cite{panagopoulos,pronin} and $\lambda(T)\sim T^{2.7}$ \cite{chen}], which is at odds with tunneling and other
measurements which indicate that there is a sizeable $s$-wave gap. Here we present the first high resolution
measurements of $\lambda(T)$ in both single crystal and polycrystalline samples of MgB$_2$. We find strong evidence for
a predominately exponential temperature dependence of $\lambda$ at low temperature consistent with $s$-wave behavior.
The gap deduced from fits to the data however, is significantly smaller than the BCS weak coupling value.

Single crystal samples of MgB$_2$ were prepared by a high pressure synthesis route as described in Ref.\ \cite{lee}.
The data presented here was taken on a plate-like crystal with dimensions 0.35 $\times$ 0.22 $\times$ 0.1 mm$^3$, the
smallest dimension being the $c$-axis. A second sample showed essentially identical behavior. The crystal orientation
was verified with an x-ray Laue camera to within $\sim 3^\circ$.  The as-grown samples are shiny and gold in color but
tarnish quickly in air.  To ensure we measured clean surfaces we immersed the crystals briefly in a $\sim$ 0.5 \%
solution of HCl in ethanol, which removed this surface discoloration.

Measurements were also made on polycrystalline MgB$_2$ samples, which were fabricated from commercially available (Alfa
Aesar) powder. This powder has a relatively wide range of grain sizes which complicates quantitative analysis of the
data.  For the analysis to give a reliable estimate of the absolute values of $\lambda(T)$ grain sizes $\lesssim 5
\lambda$ are required \cite{porch93}.  To obtain such a distribution the powder was ground in an agate mortar and then
sedimented in acetone for one hour\cite{athanassopoulou}.  The resulting powder was then cast in epoxy ($\sim$6\% by
volume). The grain size distribution was measured from scanning electron microscope images. The diameter of 96\% of the
grains was less than 1$\mu$m, and the mean diameter was 0.56$\mu$m.

Measurements of penetration depth were performed in a tunnel diode oscillator operating at 11.9~MHz\cite{carrington99},
with frequency stability of a few parts in $10^{10}$~Hz$^{-\frac{1}{2}}$.  This translates to a resolution in $\lambda$
of $10^{-12}$m for our powder samples and $10^{-10}$m for the single crystals. A particular feature of our apparatus is
the very low value of the ac-probe field which we estimate to be $\sim 1 \mu$T. Ambient dc fields are shielded to a
similar level with a mu-metal can. Changes in the oscillator frequency are directly proportional to the inductance of
the probe coil and hence to the susceptibility of the sample.  For single crystal samples, this frequency shift can be
directly related to changes in the penetration depth using the known sample dimensions \cite{prozorov00}. For
polycrystalline samples the relation between the measured susceptibility (per unit volume of superconductor) $\chi$ and
$\lambda$ is more complicated and depends on the size distribution of the grains.  For well separated grains
\begin{equation}
\chi = \frac{-\frac{3}{2}\sum\limits_i \left(1-
\frac{3\lambda}{r_i}\coth(\frac{r_i}{\lambda})+\frac{3\lambda^2}{r_i^2}\right)r_i^3 N_i }{\sum\limits_i r_i^3 N_i }
\label{chieq}
\end{equation}
where $N_i$ is the measured number of grains of radius $r_i$ \cite{porch93} which are assumed to be spherical
\cite{waldram}. $\lambda(T)$ of our polycrystalline sample was determined from the measured $\chi(T)$ and grain size
distribution by solving Eq.\ (\ref{chieq}) at each temperature point.

\begin{figure}[b]
\centerline{\psfig{figure=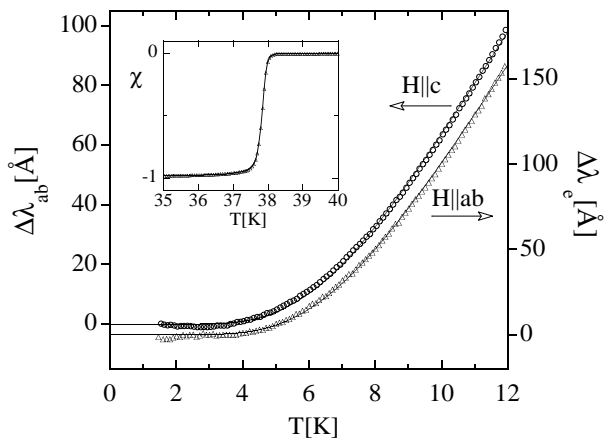,width=8cm}} \vskip 0cm \caption{Temperature dependence of the penetration depth
for an MgB$_2$ single crystal. Data for $H$ parallel to $c$ (left axis) and $ab$ (right axis) are shown.  The solid
lines are fits to Eq.\ (\ref{bcsfit}). The inset shows the susceptibility near $T_c$.} \label{xtllamfig}
\end{figure}

The superconducting transition of the single crystal is shown in the inset to Fig.\ \ref{xtllamfig}.  The $T_c$ was
38.0~K and the 10-90\% transition width was 0.3~K showing that the crystal was of high quality. The main part of this
figure shows the low temperature ($T<$12~K) behavior of $\lambda(T)$ for this crystal with the field applied along each
of the two principle crystallographic directions.  $\Delta \lambda (T)$ denotes the change in $\lambda$ relative to
that at our base temperature ($T$=1.35~K). The striking feature of these curves is the lack of temperature dependence
of $\Delta \lambda(T)$  below around 4~K, which points to the existence of an energy gap in all directions in
$k$-space.

For $H\|c$ only in-plane currents are excited and the in-plane penetration depth ($\lambda_{ab}$) is measured directly,
within a calibration factor which depends on the dimensions of the crystal. This calibration factor was estimated using
the procedure outlined in Ref.\ \cite{prozorov00}.  As the faces of the crystal were somewhat uneven we estimate that
the absolute values of $\Delta \lambda (T)$ are only accurate to about 20\%. For $H\|ab$ both in-plane and out-of-plane
currents screen the field and so we measure a mixture of $\lambda_{ab}$ and the out-of-plane penetration depth
$\lambda_c$.  As $\lambda\ll l_a,l_b,l_c$ ($l_a,l_b,l_c$ are the dimensions of the crystal along the $a,b,$ and $c$
directions respectively), the effective volume of the crystal penetrated by the field is approximately (for $H\|a$) $V
= 2\alpha(l_al_b\lambda_b +l_al_c\lambda_c)$, where $\alpha$ is the field enhancement due to the demagnetising effects
($\alpha=1.2$ for $H\|{ab}$). In the figure we show an effective $\lambda$, $\Delta \lambda_e = \Delta V /(2 \alpha
l_al_b) = \Delta \lambda_b + l_c/l_b\Delta\lambda_c$, which would be equal to $\Delta\lambda_b$ if the crystal were
thin. Again there is some uncertainty in the absolute values ($\sim$ 20\%) because of the uneven surfaces. For this
reason we do not attempt to determine $\Delta\lambda_c$ from these data. However, we note that $\Delta \lambda_e(T)$
has a very similar $T$ dependence to $\Delta \lambda_{ab}$ but is approximately 1.4 times larger. With the aspect ratio
of the crystal ($l_c/l_b$) equal to 2.2, we conclude that $\Delta \lambda_c$ is between 1.5 and 2.5 times larger than
$\Delta \lambda_{ab}$ up to 12~K.

For our crystals $\lambda_{ab}(0) \simeq 1100$~\AA, $\xi_{ab}(0)\simeq $55 \AA \cite{lee}, and the mean free path
(determined from the resistivity at 40~K) is $\sim$ 250~\AA \cite{lee,canfield}.  We therefore expect to be in the
moderately clean, local limit. Hence, for $s$-wave pairing we expect, for $T\lesssim T_c/3$, that $\lambda(T)$ should
follow the BCS behavior
\begin{equation}
 \Delta \lambda(T) \simeq \lambda_0 \sqrt {\frac{\pi \Delta_0}{2T}} \exp \left( -\frac{\Delta_0} {T}\right)
\label{bcsfit}
\end{equation}
were $\Delta_0$ is the value of the energy gap at zero temperature and $\lambda_0=\lambda(0)$ in the weak coupling
case.  In Fig.\ \ref{xtllamfig} we show a fit of Eq.\ (\ref{bcsfit}) to the crystal data in the two orientations.  The
fit to $\Delta\lambda_{ab}$ ($H\|c$) gives $\Delta_0 = 29\pm 2$~K, and to $\Delta\lambda_e$ ($H\|ab$) gives $\Delta_0 =
32\pm 2$~K.  We note the remarkable similarity between these two values (the error reflects the different values of
$\Delta_0$ obtained as the upper limit of the fit was varied from 10~K to 15~K).

\begin{figure}[b]
\centerline{\psfig{figure=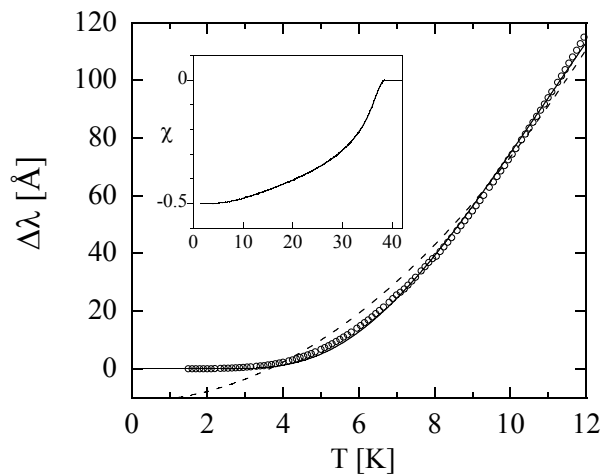,width=8cm}} \vskip 0cm \caption{Temperature dependence of the penetration depth
for a sedimented polycrystalline sample of MgB$_2$. Fits to the BCS expression [Eq.\ (\ref{bcsfit})] (solid line) and a
T$^2$ law (dashed line) are shown. The inset shows the susceptibility over the full temperature range.}
\label{polylamfig}
\end{figure}

In Fig.\ \ref{polylamfig} we show $\lambda(T)$ calculated from the susceptibility of the sedimented polycrystalline
sample. The data are very similar to that of the single crystals.  A fit of the data to Eq.\ (\ref{bcsfit}) gives
$\Delta_0 = 30\pm 2$~K, which is in very good agreement with the values found for the single crystal.  The
susceptibility ($\chi$) over the full temperature range is shown in the inset. The lack of sharp change in $\chi$ at
$T_c$ for this sample, is entirely due to the small size of the grains [compared to $\lambda(T)$] obtained by the
sedimentation procedure and does not indicate inhomogeneity. This is confirmed by the calculated temperature dependence
of the superfluid density (see below) and the sharp transition observed in an unsedimented sample (not shown).

As mentioned above, several other authors \cite{panagopoulos,chen,pronin} have claimed that $\lambda(T)$ in MgB$_2$
does not follow a simple exponential $T$ dependence but rather a power law dependence with an exponent close to 2. In
Fig.\ \ref{polylamfig} we show a $T^2$ fit to the polycrystalline data. Clearly, this fit is much worse than the BCS
dependence. We have also tried to fit other forms such as $AT+BT^2$ or $AT^2/(T+T^*)$ but no significant improvement
was found.  The same conclusion is reached from fits to the single crystal data. The key reason for this difference is
that previous studies have been limited to temperatures above $\sim 4$~K, whereas $\Delta\lambda(T)$ only shows a clear
signature of gapped behavior below this temperature.

\begin{figure}[b]
\centerline{\psfig{figure=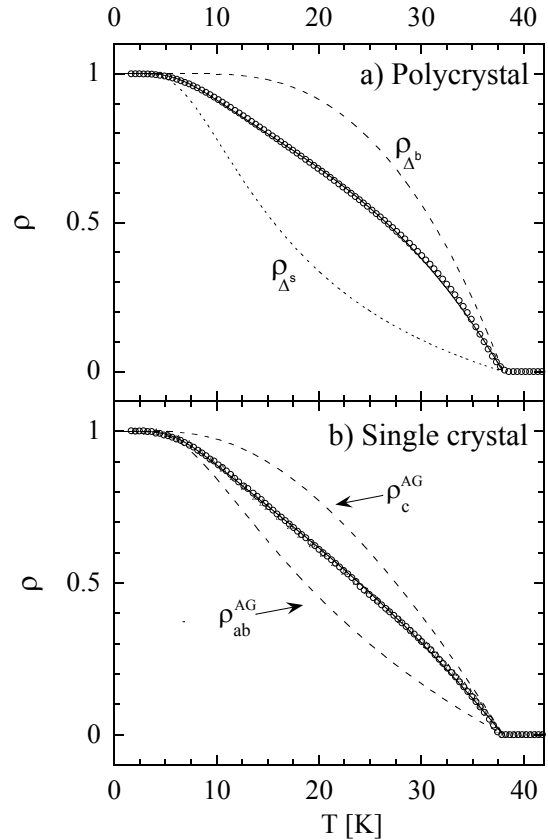,width=7.5cm}} \vskip 0cm \caption{(a) Superfluid density
[$\rho=\lambda^2(0)/\lambda^2(T)$] of the polycrystalline sample ($\circ$), along with a fit to the multigap model
(solid line).  The contribution of the small gap ($\rho_{\Delta^s}$) and the large gap ($\rho_{\Delta^b}$) in the model
are also shown. (b) In-plane superfluid densities
 of the single crystal sample [$\rho_{ab}$ ($\circ$) and $\rho_e$ ({\tiny $\triangle$})],
 along with a fit to the multigap model (solid line).  The predicted behavior of $\rho_{ab}$ and $\rho_c$ in
 the anistropic gap model for $a=2.2$ is shown by the dashed lines (denoted $\rho^{AG}_{ab}$ and $\rho^{AG}_c$).}
\label{rhofig}
\end{figure}

From the polycrystalline data we are able to deduce absolute values of $\lambda(T)$.  We find that
$\lambda(0)=1600\pm200$~\AA~ which is in good agreement with other studies \cite{chen,finnemore}.  In Fig.\
\ref{rhofig}a we use this value to calculate the superfluid density $\rho=[\lambda(0)/\lambda(T)]^2$ for this sample.
This uncertainty in $\lambda(0)$ does not make any significant difference to the temperature dependence of $\rho(T)$.

MgB$_2$ has an anisotropic structure, with Mg atoms sandwiched between planar, hexagonal boron rings.  It is therefore
expected that there will be some anisotropy between the $ab$-plane and the $c$-axis responses.  Recent results on
aligned crystallites and single crystals have shown a significant anisotropy in $H_{c2}$ which implies an anisotropy in
the coherence lengths, $\gamma=\xi_{ab}/\xi_{c}$.  In an anisotropic Ginzburg-Landau theory this implies a similar
anisotropy in $\lambda$, $\gamma=\lambda_c/\lambda_{ab}$. There is some disagreement about the magnitude of this
anisotropy, Ref.\ \cite{lima} gives $\gamma=1.7\pm0.1$ and Refs.\ \cite{xu,lee} give $\gamma=2.6\pm0.1$. We note that
the latter of these measurements were conducted on crystals identical to the ones measured here. As far as we know
there is no general solution for the moment of a sphere when $\lambda$ is anisotropic, however solutions do exist in
the limits $\lambda \gg r$ \cite{kogan} and $\lambda \ll r$ \cite{kufaev}. Fortunately, these two limits give similar
results for $\gamma\lesssim 3$. To within $\pm$10\% the effective $\lambda(0)$ equals 1.2 $\lambda_{ab}$ or 1.5
$\lambda_{ab}$ for $\gamma$=1.6 and 2.6 respectively.  We can use this fact to estimate the values of $\lambda_{ab}(0)$
and $\lambda_c(0)$ from our polycrystalline data. We find that $\lambda_{ab}$=1300~\AA~ and $\lambda_c$=2100~\AA~ or
$\lambda_{ab}\simeq 1100$ ~\AA~ and $\lambda_c\simeq 2800$ ~\AA~, for the two values of $\gamma$ respectively.  A value
of $\lambda_c$ roughly two times $\lambda_{ab}$ is consistent with the much stronger $T$ dependence of $\Delta\lambda$
in the $H\|ab$ configuration as discussed above.

Using the value of $\lambda_{ab}=1100$~\AA~(appropriate to the anisotropy of our crystals), we calculated the in-plane
superfluid density (Fig.\ \ref{rhofig}b) from the single crystal data for $\Delta\lambda_{ab}(T)$.  The overall
dependence is similar to that of the polycrystalline data, suggesting that there is not a significant anisotropy in the
temperature dependence of $\rho$. Although the $H\|ab$ data ($\Delta\lambda_e$) is a mixture of $\lambda_{c}$ and
$\lambda_{ab}$, we note that if we choose $\lambda_e(0)=1750$~\AA~the calculated superfluid density
($\rho_e=\lambda^2_e(0)/\lambda^2_e(T)$) almost exactly overlaps $\rho_{ab}(T)$, showing that the two differ by only a
simple scale factor  and supporting the conclusion that there is little anisotropy in the temperature dependence of
$\rho$.

Although the overall temperature dependence of $\lambda$ deduced from our measurements is consistent with an $s$-wave
BCS picture, the value of the gap obtained is considerably smaller than the usual BCS weak coupling value
($\Delta_0/T_c= 1.76$).  Similar conclusions have been reached from some tunneling data \cite{bollinger}, and specific
heat studies \cite{bouquet}.

The calculated band structure of MgB$_2$ is composed of two sets of distinct bands; quasi-two-dimensional tubes which
run along the $c$ axis and a roughly three-dimensional network \cite{kortus}.  Several authors have proposed that there
may be different gaps associated with these two bands \cite{bouquet,liu}.  For these band specific gaps to remain
distinct the scattering between bands needs to be weak, and there needs to be some weak coupling between the bands so
that the $T_c$s are the same. Bouquet {\it et al.}\cite{bouquet} have recently proposed a simple phenomenological model
to calculate the thermodynamic properties in such a two band system.  The model is based on the so-called $\alpha$
model which is often used to describe strong coupling effects\cite{padamsee}.   It is supposed that the gaps on the two
bands $\Delta^b$ and $\Delta^s$ follow the usual BCS $T$ dependence, but have $T=0$ values which differ from the BCS
one. The superfluid densities from each band add together to give the measured total.  These two gaps and the ratio of
the contributions of each to the total superfluid density are fitting parameters.

In Fig.\ \ref{rhofig} we show a least squares fit of both the polycrystalline and single crystal data to this model.
The agreement with the data is extremely good in both cases. We find that for the polycrystalline data,
$\Delta^s=30$~K, $\Delta^b=89$~K with relative proportion 40:60 respectively. For the in-plane single crystal data the
parameters are found to be: $\Delta^s=29$~K, $\Delta^b=75$~K with relative proportion 45:55.  The small gap
($\Delta^s$) is identical to that found from the fits of $\Delta \lambda(T)$ to Eq.\ (\ref{bcsfit}). Bouquet {\it et
al.} have also applied this model to heat capacity data and find very similar results for both the gap values and their
relative weights.

First principles, density functional theory calculations of the electron phonon coupling in MgB$_2$, have been
performed by Liu {\it et al.}\cite{liu}.  These calculations predict the existence of two distinct gaps with almost
equal weights ; a small gap ($\Delta/T_c=0.65$) associated with the 3D sheets and a larger gap ($\Delta/T_c=2.0$)
associated with the 2D tubes.  Both the size of the gaps and their relative weights are in good agreement with the
Bouquet model.  In addition, because the smaller gap is located on the 3D sheet, we expect the effective gap, derived
from fits to $\Delta \lambda(T)$ at low temperature to be fairly isotropic. This is also in agreement with our
observations.   We note however, that Raman scattering measurements \cite{quilty} on crystals prepared identically to
those studied here, have clearly identified only one sharp pair-breaking peak at 105 cm$^{-1}$ corresponding to a gap
value $\Delta(0)/T_c=2.0$.  Although there is some scattering below this peak, a second peak corresponding to a second
smaller gap has not been seen.

An alternative model \cite{haas} assumes that there is a single $s$-wave gap which is anisotropic, having a $k$
dependence, $\Delta(z)=\Delta (1+a z^2)/(1+a)$, where $z$ is the cosine of the polar angle and $a$ is a fitting
parameter which controls the anisotropy.  Following the procedure outlined in Ref.\ \cite{haas} we have calculated the
superfluid density in this model for arbitrary $a$.  We find that to fit the observed ratio of $\Delta/T_c = 0.75$ we
need $a=2.2\pm0.4$, or a total gap anisotropy $(1+a) = 3.2\pm0.4$. The overall $T$-dependence of $\rho_{ab}$ calculated
for this value of anisotropy is shown in Fig.\ \ref{rhofig}b and can be seen to be in serious disagreement with the
data.  Better agreement can be found if we set $\lambda_{ab}(0)\sim 650$~\AA. However, this is almost a factor 2 below
our estimate and well outside our expected error.  The very different behavior of $\rho_c$ in this model (see figure)
is also at odds with observed correspondence between $\rho_{ab}$ and $\rho_e$.  Our data therefore clearly favor the
multigap model above.

In conclusion we have presented the first study of the penetration depth of both single crystal and polycrystalline
MgB$_2$ samples. In agreement with other direct probes of the symmetry of the order parameter we find that $\lambda(T)$
is well described by an $s$-wave behavior, but with a minimum gap that is significantly less than the weak coupling BCS
value. Of the various explanations for the small gap value we find that the two-gap model of Bouquet {\it et al.} is in
best agreement with the data.  We conclude therefore, that although the pairing interaction in this compound may be
phononic, the gap structure is far from conventional.

Useful discussions with R.W.\ Giannetta, R.\ Prozorov, F.\ Bouquet, K.\ Maki, P.\ Timms, J.F.\ Annett and J.R.\ Cooper
are acknowledged.  This work was partially supported by the New Energy and Industrial Technology Development
Organization as collaborative research and development of fundamental technologies for superconductivity applications.

\end{document}